\documentclass[%
 reprint,
 amsmath,amssymb,
 aps,
 prb, 
 floats,
]{revtex4-1}

\usepackage{graphicx}
\usepackage{dcolumn}
\usepackage{bm}

\begin{document}

\preprint{}

\title{Fundamental Study of Hydrogen Segregation at Vacancy and Grain Boundary in Palladium}

\author{Hieu H. Pham}
 \email{Current address: Lawrence Berkeley National Laboratory (hhpham@lbl.gov)}
\author{Tahir {\c C}a{\u g}{\i}n}
 \email{Corresponding author: T.C. (tcagin@tamu.edu)}
\affiliation{ Artie McFerrin Department of Chemical Engineering
\\Texas A\&M University, College Station, TX 77845}


\begin{abstract}
We have studied the fundamental process of hydrogen binding at interstitial, vacancy and grain boundary (GB) in palladium crystals using Density-Functional Theory. It showed that hydrogen prefers to occupy the octahedral interstitial site in Pd matrix, however a stable H-vacancy complex with most H occupations would contain up to eight hydrogen atoms surrounding the vacancy at tetrahedral sites. Furthermore, H presence assists the pairing or formation of nearby vacancies, which in agreement with previous suggestions by both experiment and theory investigation. Also, this observation could imply about a hydrogen embrittlement (HE) mechanism through the connections of microvoid and cracks. The segregation of hydrogen at grain boundary, nevertheless, has shown a different effect. High H accumulation results in grain boundary extension, which is related the HE mechanism of grain decohesion observed by experiments.
\end{abstract}

\pacs{Valid PACS appear here}

\maketitle

\section{\label{intro} Introduction}

The absorption of hydrogen in metals, in particular Pd-based materials,  has attracted great interest due to its unique behavior, including its high solubility and activity~\cite{ColdFusion, Devanathan62, Conrad1974435, Favier21092001}. However, it is also known that the interaction of impurity and the defects in host materials would drastically affect their properties and performance. Specifically, both experimental and theoretical investigations have evidenced that defects such as vacancy, dislocations and grain boundary in metals can attract a high localization of impurities,~\cite{Birnbaum1994191, Fukai2000121, Fukai73.1640, Pham20105142, arXiv150507524P} and subsequently are responsible for many prominent phenomena.


These effects, nevertheless, require a threshold concentration of hydrogen at intrinsic binding sites. Therefore, the learning of the H saturation at these locations could be critical for the understanding of their behavior. In general, hydrogen diffusion in crystals is active by first occupying interstitial positions until they are trapped in open volumes (such as vacancy or grain boundary). In fact, it was suggested that the coalescence of vacancies in the presence of H will provide source for microcrack initiation and subsequent embrittlement~\cite{LuKaxiras94.155501}. Also, experiments performed by Fukai et al. evidenced that high H absorption can introduce the formation of superabundant vacancies in several transition metals ~\cite{Fukai2000121, Fukai73.1640}. Grain boundaries, on the other hand, can serve as ideal transport channels for impurity elements, at the same time could also be prefered nucleation site of dislocation and voids that results in stress corrosion cracking.~\cite{Kart20091236, PhamJAP} Some authors suggested that hydrogen causes embrittlement in Pd by decreasing the cohesive strength of GB, since H acts electron-receptor and weaken the bonding of host atoms across GB.~\cite{Bond19891407, Liu199225}


Due to the intricate nature of these phenomena, real-time measurements and observations can be restricted. ~\cite{Myers64.559} Recent advances in computer powers and  computational theory have made the atomistic simulation an excellent tool for the investigation of materials properties. In fact, first-principles calculations have been widely used and succeeded in describing structural properties at electronic level, particularly for the treatment of defect phenomena in solids. ~\cite{Rice198923, Wu15071994, Yamaguchi21012005, Geng2001585, Geng63.165415, PhamPCCP2, Kart20091236} In this paper, we have examined the interactions of hydrogen with defects in both Pd single crystal and bicrystal. The segregation of H at grain boundary and vacancy up to their saturated state was investigated in order to understand the process of H localization in palladium at the very fundamental level.

\section{\label{method} Computational method}

The Density Functional Theory (DFT),~\cite{Kohn140.A1133} as implemented in Vienna Ab-initio Simulation Package (VASP),~\cite{Kresse54.11169} was employed to perform the first-principles calculations. In this work, the crystal properties of Pd were described by the electron Projector-Augmented Wave (PAW) methods~\cite{Bloch50.17953} with the PBE Generalized Gradient
Approximation (GGA) exchange-correlation, ~\cite{PBE77.3865} unless noted otherwise. Bulk Pd crystal was simulated using a supercell of 108 face-centered cubic (FCC) atoms; in which the convergence of total energy is 0.1 meV per atom. Additionally, a bicrystal model of 80 atomic sites was constructed to simulate a symmetric tilt $\Sigma$5/(210)/$\langle100\rangle$ grain boundary in Pd. In the k-space sampling, we have used a Monkhorst-Pack grid~\cite{KPOINTS13.5188} of 4$\times$4$\times$4 for the single-crystal and 4$\times$4$\times$2 for bicrystal supercells, for the plane wave basis, respectively. In calculating the total energy, equilibrated structures are obtained by relaxing the atomic positions and the lattice vectors until stress components $\sigma$$_{xx}$, $\sigma$$_{yy}$ and $\sigma$$_{zz}$ are all well below 1.0 kbar. Periodic boundary conditions are imposed in three dimensions.

The precipitation of impurity atoms ensues from the migration through grain boundaries, as well as the generation of vacancies. To demonstrate the binding tendency between hydrogen impurity and palladium host atoms, the concept of absorption energy is used; its average value, E$_{abs}$$^{avg}$, from absorption of n hydrogen atoms, is calculated by the following equation:~\cite{Yamaguchi21012005}

\begin{equation}\label{absorption}
 nE_{abs}^{avg}[nH] = E_{tot}^{nH} - E_{tot}^{0H} - \frac{n}{2}E^{H_{2}}
\end{equation}
where E$_{tot}$$^{nH}$ and E$_{tot}$$^{0H}$ describe total energy of systems with n hydrogen segregation and at clean state, respectively. E$^{H_{2}}$ is the total energy of one isolated hydrogen molecule, calculated by placing that molecule in a large supercell (20$\times$20$\times$20 {\AA}).

The formation energy of one vacancy in any system is obtained from the change in total energies of that system before ($E_{tot}$) and after the formation of vacancy (E$_{tot}$$^{v}$), adding the average energy of one Pd atom in bulk state (E$_{bulk}$$^{Pd}$):
\begin{equation}\label{vacancy}
 E_{f} = E_{tot}^{v} - E_{tot} + E_{bulk}^{Pd}
\end{equation}

In order to illustrate the site competition between H segregation in vacancy (or grain boundary) with H binding in interstitial, we calculate the trapping energy of H. Its average value therefore is calculated with respect to the change in total energy of bulk palladium E$_{tot}$$^{1H}$-E$_{tot}$$^{0H}$, by adding one H as interstitial:~\cite{Vekilova80.02410}

\begin{eqnarray}
  E_{tr}^{vac}[nH] = \frac{1}{n}(E_{tot}^{nH-v} - E_{tot}^{0H-v}) - (E_{tot}^{1H} - E_{tot}^{0H}) \\
  E_{tr}^{gb}[nH] = \frac{1}{n}(E_{tot}^{nH-gb} - E_{tot}^{0H-gb}) - (E_{tot}^{1H} - E_{tot}^{0H}) \label{trap_avg}
\end{eqnarray}
where E$_{tot}$$^{xH-v}$ (or E$_{tot}$$^{xH-gb}$) denotes the total energy when we have a complex of x H atoms segregated at a vacancy (or grain boundary) in the system. The trapping energy of the single n$^{th}$ H atom can be estimated by the difference in total trapping energies between n and (n-1) H atoms:

\begin{eqnarray}\label{trap_n}
  \Delta E_{tr}^{vac}[n] = nE_{tr}^{vac}[nH] - (n-1)E_{tr}^{vac}[(n-1)H]\\
  \Delta E_{tr}^{gb}[n]  = nE_{tr}^{gb}[nH]  - (n-1)E_{tr}^{gb}[(n-1)H]
\end{eqnarray}

\section{\label{result} Results and discussion}
 \subsubsection{\label{point} Hydrogen binding at interstitial and vacancy}

\begin{figure}[!t]
\includegraphics[width=6 cm]{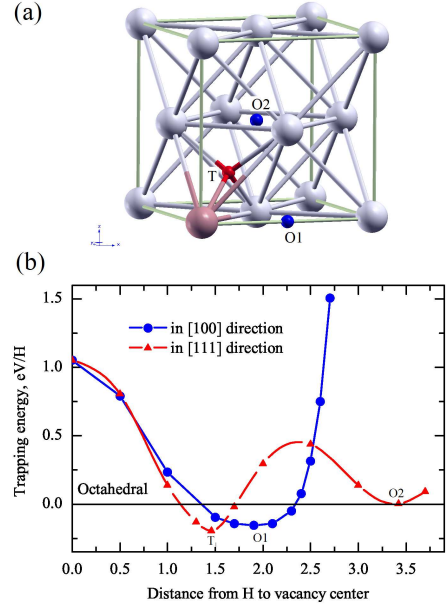}
\caption{\label{fig1} (color online) (a) Atomic configuration of a palladium face-centered cubic unit cell, showing tetrahedral (red) and octahedral (blue) interstices. In perfect FCC Pd, octahedral site is more energetically favorable for H binding than tetrahedral H binding. (b) Variation of trapping energy of H along diffusion paths into the vacancy (darker Pd position in the figure above). The zero level corresponds to the binding energy of H to octahedral interstitial. }
\end{figure}

For a simple crystal such as FCC palladium, octahedral (O-site) and tetrahedral (T-site) are two common types of interstices. The nearest tetrahedral interstitial (T) is in [1 1 1] direction. The nearest octahedral interstitial (O1) is in [1 0 0] and the second nearest one (O2) is in [1 1 1] as at the center of the unit cell body (Fig.~\ref{fig1}a). Using equation~\ref{absorption}, the calculations of bulk Pd show that the binding (or absorption) energy of H to O-site is -0.1 eV, which is stronger than that at T-site (-0.05 eV). This suggests a preference of H occupation at O-site over the other in Pd matrix.~\cite{Dietrich79} However, as the magnitude of this binding energy is quite low, the stability of H at interstitial sites is subject to other factors. 

It is believed that hydrogen binding could be stronger at other defects such as vacancy and grain boundary. In Fig.~\ref{fig1}b we present the variation of H trapping energy along diffusion paths into the vacancy (dark Pd sphere). Trapping energy is calculated with respect to binding energy of one H atom to the octahedral interstitial (the zero level corresponds to the binding energy of H to octahedral interstitial). In agreement with previous works in metals,~\cite{Vekilova80.02410, LuKaxiras94.155501} our calculation also showed that the center of vacancy is a very unfavorable binding site for H. Parts of the curves below the zero level indicate possible regions around a vacancy at which H binding is stronger than that of a general octahedral binding (with no vacancy). At the same time, this indicates that the formation of the nearest-neighbor vacancy makes H binding at octahedral site stronger (O1 in Fig.~\ref{fig1}) . However, it is interesting that the most energetically stable site for H-vacancy binding is in [111] direction, close to its tetrahedral position (T in Fig.~\ref{fig1}) . This H bonding will be threefold, instead of fourfold, due to one missing Pd (vacancy). Farther in [111] direction, vacancy has no effect on H binding at its second nearest octahedral interstitial (as trapping energy at O2 site is zero).

\begin{figure}[!b]
\includegraphics[width=6 cm]{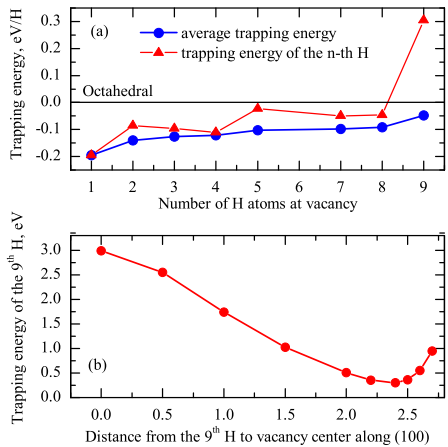}
\caption{\label{fig2} (color online) (a) Variation of H trapping energy  as a function of H number at vacancy site. A Pd-H configuration with hydrogen filling up 8 tetrahedral sites around vacancy is energetically favorable and structurally stable. 
(b) Trapping energy of the 9$^{th}$ H as function of distance in [100] from vacancy center. Its values is everywhere positive along [100] direction}
\end{figure}

Using local-density approximation (LDA) method, Vekilova et. al~\cite{Vekilova80.02410} reported about a possibility of multiple hydrogen occupancy, in which a maximum of six hydrogen atoms can be trapped in a monovacancy. The most favorable sites were reported to be along the directions of [100] family, as six octahedral sites coordinate a Pd atom. However, in our work with GGA, that configuration of VacH$_{6}$ is not energetically favorable. Instead, we found that the stable structure of monovacancy with most H will contain eight H atoms,  surrounding the vacancy at its tetrahedral sites. With this maximum number of H occupations, the H binding is still competitively stronger than the general octahedral binding in bulk (Fig.~\ref{fig2}a). However, the insertion of any extra H into this VacH$_{8}$ configuration will result in a positive trapping energy, i.e. the extra H is likely to diffuse away and occupy an interstitial site elsewhere.  Fig.~\ref{fig2}b shows the variation of trapping energy of the 9$^{th}$ H with respect to its distance from vacancy center in [100] direction. There is nowhere to fit the 9$^{th}$ H into this VacH$_{8}$ complex to make it a stable complex.

\begin{table}[]
\caption{\label{table1}
Formation energy of monovacancy, divacancy and influence of hydrogen segregation
}
\begin{ruledtabular}
\begin{tabular}{lcdr}
\textrm{Vacancy type}&
\textrm{Formation energy E$_{f}$, eV}\\
\colrule
Mono-vacancy & 1.17 \\
Mono-vacancy expr.\footnotemark[1]
& 1.50 \\
Second vacancy in [100] of VacH$_{0}$ & 1.19 \\
Second vacancy in [110] of VacH$_{0}$ & 1.13 \\
Second vacancy in [100] of VacH$_{8}$ & 1.22 \\
Second vacancy in [110] of VacH$_{8}$ & 1.04\\
\end{tabular}
\end{ruledtabular}
\footnotetext[1]{Ref.~\onlinecite{Kraftmakher199879}}
\end{table}

In addition, the vacancy formation process can be affected by different factors, including the presence of neighboring vacancies and impurity atoms. It is commonly believed that the formation of a di-vacancy is less costly than the formation of two separate vacancies. However, our calculations show that this fact is observed only in [110] direction, i.e. di-vacancy formed by the pairing of two nearest neighbors (Table~\ref{table1}). The formation of the second vacancy in [100] is shown to bear little to no effect.

 \subsubsection{\label{gb} Hydrogen segregation at grain boundary}

\begin{figure}[!b]
\includegraphics[width=6 cm]{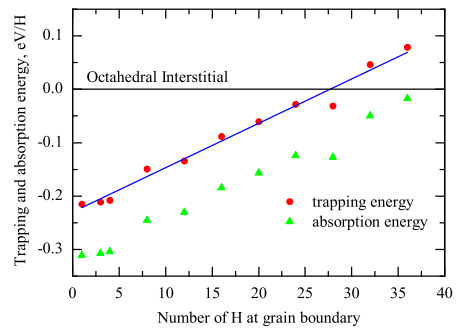}
\caption{\label{fig3} (color online) Average trapping energy as a function of H number at grain boundary region. The absorption energy differ from trapping energy by the value of E$_{b}$ at octahedral site in bulk.
}
\end{figure}

\begin{figure}[!t]
\includegraphics[width=8 cm]{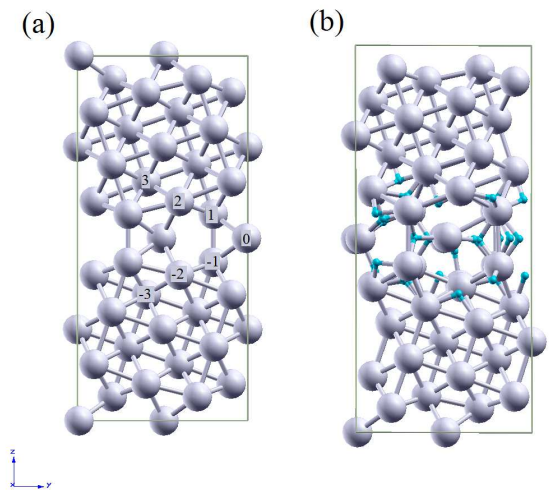}
\caption{\label{fig4} (color online) Atomic configurations of Pd bicrystals at H-clean state (a) and with saturated segregation of 28 H atoms. Large spheres demonstrate palladium and small ones are hydrogen atoms. Trapping energy of H decreases with increase in its local concentration at grain boundary. When H local concentration at grain boundary  exceeds the saturation level, hydrogen is attracted by interstitial sites in bulk.}
\end{figure}

In this subsection, we will discuss the occupation of H at a $\Sigma$5 (210) grain boundary, as mentioned earlier that grain boundaries play a significant role in the issue of impurity transport and segregation. Using equation~\ref{trap_avg}, the average trapping energy was calculated as a function of H occupation at this GB (Fig.~\ref{fig3}). The absorption (segregation) energy differs from trapping energy by the value of one H octahedral binding. According to Rice-Wang theory,~\cite{Rice198923}, it was commonly accepted that H is an interfacial embrittler by segregation energy.~\cite{GengOlso2005}

Our simulation showed that the average segregation energy is negative for H occupations up to 36 atoms at this $\Sigma$5 (210) grain boundary. In other words, a complex of GBH$_{36}$ could be thermodynamically feasible. However, Fig.~\ref{fig3} indicated that the average trapping energy is positive with approximately more than 28 H occupations, , i.e. further absorption of H is energetically unfavorable.
Fig.~\ref{fig4} shows the atomic configuration of Pd bicrystal cells at H-clean state and with 28 H occupation around GB plane. If only five Pd atom layers around GB are taken into account, then the local H concentration is roughly 140 \% at. (Pd$_{20}$H$_{28}$)

Due to the geometry of GB, there are hollow sites on GB plane that could host impurity atoms. However, H atoms are unlikely to stay in those empty spaces or form hydrogen gas themselves; instead, they bind closely with Pd. The average H-Pd bond length is approximately 1.8 {\AA}, which is roughly equivalent to a bond between a tetrahedral H and Pd in bulk. Also, we found that the most common H-Pd bonding at GB is threefold. This configuration contains an H atom bound with Pd at three vertices of a (quasi-)tetrahedron, while the the forth vertex is missing. This H-Pd binding is identical to the configuration of H-vacancy complex, while H found itself most comfortable at the tetrahedral interstitial (but bonds with only three Pd due to one missing Pd vacancy, Fig.~\ref{fig1}a). While H binds threefold, it displays that one Pd atom can also participate in several H-Pd bonds (Fig.~\ref{fig4}b).

To differentiate between Pd atom layers at grain boundary, we index them GB0, GB1, GB2 ..., depending on how far they are from the grain boundary plane (Fig.~\ref{fig4}a). Due to the symmetry characterization of this GB, layers with opposite indices are equivalent (e.g. GB1 and GB-1). When H atoms surround Pd, firstly, they weaken Pd-Pd bond. Secondly, Pd-Pd can even be broken, as there is an expansion between Pd layers due to high H-GB occupancy. For instance, the distance between layers GB2 and GB-2 increase from 3.83 {\AA} (at H-clean state) up to 4.82 {\AA} at the suggested H saturation (28 H atoms for this 80-Pd bicrystal supercell).

Next, we investigate the question of how H localization  would influence the formation of vacancy around the GB domain. As presented in Table ~\ref{table2}, the first Pd layers (GB1 and GB-1) are more likely to form vacancy, with the formation energy significantly reduced. Farther away, the effect of grain boundary and vacancy formation would be minimal.
The presence of hydrogen in this case, however, has minimal effect on the formation of vacancy at the GB vicinities. For instance, with eight (or twenty-eight) H atoms segregation at grain boundary, the vacancy formation energy at GB1 even increases slightly to 0.55 eV (or 0.63 eV, repectively), compared to its value of 0.48 eV in H-clean state. Therefore, unlike H-vacancy binding, H-GB segregation doesn't tend to initiate hydrogen embrittlement through microvoid formation. Instead, high H content could result in a decohesion across the Pd grain boundary and subsequently, the hydrogen embrittlement could be initiated by the grain separation at sufficient external stress. This observation is in agreement with calculations using Rice-Wang model, in which single H was reportedly a GB embrittler by reducing the grain cohesion.~\cite{Zhong62.13938, GengOlso2005, Kulkov2009}


\begin{table}[!t] 
\caption{\label{table2}
Formation energy of monovacancy at different sites around grain boundary, eV}
\begin{ruledtabular}
\begin{tabular}{lcdr}
\textrm{Vacancy site} &
\textrm{Formation energy E$_{f}$, eV} \\
\colrule
GB0 & 1.29 \\
GB1 & 0.48 \\
GB2 & 1.00 \\
GB3 & 1.04 \\
GB4 & 1.06 \\
GB5 & 1.09 \\
\end{tabular}
\end{ruledtabular}
\end{table}

\section{\label{conclusion} Conclusions}
We have reported the saturated localization of H binding at vacancy and grain boundary in Pd crystals. Theoretical investigations for favorable configurations of H-vacancy and H-GB complexes from low to high H occupation were presented.

While H prefers to occupy the octahedral interstitial position, they are anticipated to fill in the tetrahedral sites in complex with a Pd vacancy. Our calculation showed that one Pd vacancy is capable of storing up to eight H, instead of six as reported previously.~\cite{Vekilova80.02410} Pairing of vacancies is encouraging in [110] direction, and H presence will further assist multi-vacancy formation, which implies the hydrogen embrittlement by connections of microvoids. Numerous experimental observation and theoretical calculations have reached to the same conclusions on vacancy formation induced by hydrogen.

In both cases of open defects (vacancy and grain boundary), hydrogen prefers a threefold bonding with palladium. However, H segregation at GB results in weakening Pd-Pd bonds across GB, by inducing the grain separation. This observation supports the decohesion mechanism of hydrogen embrittlement as evidenced by many experiments.


\begin{acknowledgments}
The research here was supported by ONR. The authors would also acknowledge the Texas A\&M Supercomputing Facility for providing computing resources.
\end{acknowledgments}

\nocite{*}
\providecommand{\noopsort}[1]{}\providecommand{\singleletter}[1]{#1}%

\end{document}